\documentclass[11pt]{article}
\newcommand{\Comment}[1]{{}}
\usepackage{amsfonts,amsthm,amsmath,amssymb,slashed}
\usepackage[textwidth = 430 pt, textheight = 630 pt]{geometry}
\usepackage{color}

\Comment{\usepackage{color}
\definecolor{MyDarkBlue}{rgb}{0.15,0.15,0.45}
\usepackage[linktocpage=true]{hyperref}
\hypersetup{
colorlinks=true,
citecolor=MyDarkBlue,
linkcolor=MyDarkBlue,
urlcolor=MyDarkBlue,
pdfauthor={
Horatiu Nastase and Jacob Sonnenschein},
pdftitle={The title},
pdfsubject={hep-th}
}

\usepackage[numbers,sort&compress]{natbib}
\usepackage{hypernat}}
\usepackage{graphicx}
\usepackage{cite}

\newcommand\ignore[1]{}
\def\one{{\,\hbox{1\kern-.8mm l}}}

\def\Tr{{\rm Tr\, }}

\def\a{\alpha}\def\b{\beta}

\def\d{\partial}

\def\Tr{\mathop{\rm Tr}\nolimits}

\newcommand{\Cset}{{\,\,{{{^{_{\pmb{\mid}}}}\kern-.45em{\mathrm C}}}}}

\newcommand{\be}{\begin{equation}}
\newcommand{\bea}{\begin{eqnarray}}

\newcommand{\ee}{\end{equation}}
\newcommand{\eea}{\end{eqnarray}}

\begin{document}

\renewcommand{\thefootnote}{\fnsymbol{footnote}}

\makeatletter
\@addtoreset{equation}{section}
\makeatother
\renewcommand{\theequation}{\thesection.\arabic{equation}}

\rightline{}
\rightline{}




\begin{center}
{\LARGE \bf{\sc Lagrangian formulation, generalizations and quantization     of null Maxwell's  knots }}
\end{center} 
 \vspace{1truecm}
\thispagestyle{empty} \centerline{
{\large \bf {\sc Horatiu Nastase${}^{a}$}}\footnote{E-mail address: \Comment{\href{mailto:nastase@ift.unesp.br}}{\tt nastase@ift.unesp.br}}
{\bf{\sc and}}
{\large \bf {\sc Jacob Sonnenschein${}^{c}$}}\footnote{E-mail address: \Comment{\href{mailto:cobi@post.tau.ac.il}}{\tt cobi@post.tau.ac.il}}
                                                        }

\vspace{.5cm}


\centerline{{\it ${}^a$Instituto de F\'{i}sica Te\'{o}rica, UNESP-Universidade Estadual Paulista}} 
\centerline{{\it R. Dr. Bento T. Ferraz 271, Bl. II, Sao Paulo 01140-070, SP, Brazil}}
\vspace{.3cm}
\centerline{{\it ${}^c$School of Physics and Astronomy,}}
\centerline{{\it The Raymond and Beverly Sackler Faculty of Exact Sciences, }} \centerline{{\it Tel Aviv University, Ramat Aviv 69978, Israel}}

\vspace{1truecm}

\thispagestyle{empty}

\centerline{\sc Abstract}

\vspace{.4truecm}

\begin{center}
\begin{minipage}[c]{380pt}
{\noindent Knotted solutions to electromagnetism are investigated as an independent subsector of the theory. 
We write down  a Lagrangian and a Hamiltonian formulation of Bateman's 
construction for the knotted electromagnetic solutions. We introduce a general definition of  the null condition and generalize the 
construction of Maxwell's theory to massless  free complex  scalar, its dual two form field, and to a massless DBI scalar.   We set up the 
framework for quantizing the theory  both in a path integral approach, as well as the canonical  Dirac  method for a constrained system.  
We make several observations about  the  semi-classical quantization of systems of null configurations.  

}
\end{minipage}
\end{center}

\vspace{.5cm}

\setcounter{page}{0}
\setcounter{tocdepth}{2}

\newpage

\section{Introduction}

Electromagnetism is a free (non-selfinteracting) theory so, according to standard lore, we wouldn't expect topologically non-trivial solutions. Indeed, solitons are 
usually found in interacting theories, as in the case of water solitons, which started the field, with John Scott Russel's observation of a solitonic wave 
in a canal in Scotland. Sometimes there is a topological reason for the existence and stability of a soliton, which is the case for "kinks" in 1+1 dimensional
scalar theories, vortices in 2+1 dimensional gauge theories, or monopoles in 3+1 dimesional gauge theories, for instance. 

But the existence of a topological constraint, of a fixed topological number, turns out to be possible even in a free theory like Maxwell electromagnetism
without sources. Thus it was realized rather late that there exist solutions with a nonzero Hopf index, or "Hopfions," and the explicit solutions were 
written only in \cite{Ranada:1989wc,ranada1990knotted} by Ra\~nada, after the early work by Trautman in \cite{Trautman:1977im}. 
The standard Hopfion solution is null in the sense of the Riemann-Silberstein (RS) vector $\vec{F}=\vec{E}+i\vec{B}$, i.e. $\vec{F}^2=0$, corresponding
to $\vec{E}^2=\vec{B}^2$ and $\vec{E}\cdot \vec{B}=0$, but there are also partially null solutions, as we will explain in the following.

These solutions are also characterized by constant and nonzero "helicities" ${\cal H}_{ij}$, with $i,j=$ electric (e) or magnetic (m), which are spatial integrals
of spatial Chern-Simons terms for $\vec{E}$ and $\vec{B}$, $\int d^3x \epsilon^{ijk}A_i\d_j B_k$ ($\vec{A},\vec{B}\in(\vec{E},\vec{B})$), 
that are conserved for the null configurations ($\vec{F}^2=0$), for which one has 
$F_{\mu\nu}F^{\mu\nu}=\epsilon^{\mu\nu\rho\sigma} F_{\mu\nu}F_{\rho\sigma}=0$. 

One construction for the solutions that we will be especially interested in is Bateman's formulation \cite{Bateman:1915}, where the electric and 
magnetic fields for null solutions  are written in terms of two complex scalar fields $\a$ and $\b$. In this formulation, we can easily 
construct much more solutions, as was first shown in \cite{Kedia:2013bw,besieris2009hopf}: take any solution and apply on it a holomorphic 
transformation $\a'=f(\a,\b),\b'= g(\a,\b)$. For instance, starting from the Hopfion with ${\cal H}_{ee}={\cal H}_{mm}\neq 0$, we can obtain 
 "$(p,q)-$knotted solutions" by  the holomorphic transformations $\a\rightarrow (\a)^p,\b\rightarrow (\b)^q$.
Moreover, starting from a topologically trivial (un-knotted) solution (but still null, to be in Bateman's construction), like one with constant, equal, and 
transverse electric and magnetic fields,  we can obtain knotted ones like the Hopfion by a conformal  transformation with 
{\em complex}, rather than real, parameters, as shown in \cite{Hoyos:2015bxa}. More solutions were found in this way. Note that a plane 
electromagnetic wave is also null, and one can apply such a procedure on it as well. In \cite{Alves:2017ggb,Alves:2017zjt}, a connection 
of null electromagnetism with fluid dynamics was used in order to explore other ways of finding solutions in both theories.

One explanation for the existence and stability of the Hopfion solution is the fact that one imposes a constraint, $\vec{F}^2=0$, that makes the 
system nonlinear, and in Bateman's construction one describes the system in terms of the variables $\a$ and $\b$, though the equations of motion 
are still Maxwell's ones. We should also note that the Hopfion and related solutions are {\em time dependent}, the distribution of energy 
density coming somewhat radially from infinity towards the origin until a minimum, and then expanding again to infinity, as seen for instance in 
\cite{Hoyos:2015bxa}. It is then of interest to describe {\em only} the null electromagnetic system in Bateman's construction, and perhaps to try 
to quantize it by itself.

This is the subject of this paper. We will construct a  relativistic Lagrangian formulation for Bateman's construction, and then 
a Hamiltonian formulation. The former   can be used for a  path integral quantization and the latter for  Dirac's  canonical 
formalism for constrained systems. 
We express the null condition in a covariant way, which makes it possible to generalize the construction to other fields 
besides electromagnetism and fluids. In particular, we consider here a free   massless  complex  scalar, its  dual, a two form field, as 
well as a massless DBI scalar. 
The covariant null condition is expressed  in terms of the energy-momentum tensor as 
\be
{T^\mu}_\mu=0\;,\;\;\;{(T^n)^\mu}_\nu=0\;,
\ee
where $ {(T^n)^\mu}_\nu$ is the $n^{th}$ product  of $T^{\mu\nu}$, with two free (uncontracted) indices.

 Finally, we explore various ways to quantize the system, including path integral and Dirac quantization, 
though we find that carrying out explicitly the program is very difficult, while semiclassical quantization seems to be fail in its standard form, 
of collective coordinate quantization.

The paper is organized as follows. In section 2 we review Bateman's construction and the Hopfion solution. In section 3 we construct the 
relativistic action for Bateman's construction and the Hamiltonian formalism. In section 4 we explore the symmetries, and 
in section 5 possible generalizations 
to other systems. In section 6 we explore path integral and Dirac quantization, and in section 7 we comment on the unsuitability of collective 
coordinate quantization. In section 8 we conclude.

\section{Knotted solutions in electromagnetism}

In this section we review knotted solutions of electromagnetism, which are  solutions that have some conserved ``helicities" 
and linking number, in the construction due to  Bateman. The simplest solution is the ``Hopfion" solution. 
For more details, see  the review \cite{arrayas2016knots}.

\subsection{Electromagnetism and helicities}

Consider Maxwell electromagnetism without sources. 
In non-relativistic notation, the corresponding Maxwell's equations are
\bea
\vec{\nabla}\times \vec{E}=-\frac{\d \vec{B}}{\d t}; && \vec{\nabla}\cdot \vec{E}=0\cr
\vec{\nabla}\times \vec{B}=+\frac{\d \vec{E}}{\d t}; && \vec{\nabla}\cdot \vec{B}\;,
\eea
where we have put $c=1$.

The electric and magnetic field are, in terms of the the scalar potential $\phi$ and the vector potential $\vec{A}$, 
\be
\vec{E}=-\frac{\d}{\d t}\vec{A}-\vec{\nabla}\phi;\;\;\;\;
\vec{B}=\vec{\nabla}\times \vec{A}.
\ee
Since we have no sources, we can work in a gauge with $A_0=\phi=0$. There is the residual gauge invariance
\be
\delta\vec{A}=\vec{\nabla}\a(\vec{x})\;,
\ee
that leaves invariant the gauge condition, since $\delta A_0\equiv \delta\phi=\d_0 \a=0$. Here $\vec{A}=\vec{A}(\vec{x},t)$.

Since $\vec{\nabla}\cdot \vec{E}=0$ in the absence of sources, 
it is useful to also introduce another vector potential $\vec{C}$ for $\vec{E}$. It is just the electric-magnetic dual of $\vec{A}$, 
and is introduced in order to have a duality symmetric formulation: 
\be
\vec{E}=\vec{\nabla}\times \vec{C};\;\;\; \vec{B}=\vec{\nabla}\times \vec{A}\;,
\ee
The relation between $\vec{A}$ and $\vec{C}$ is
\be
-\frac{\d}{\d t}\vec{A}\equiv \vec{\nabla}\times \vec{C}.
\ee 
The electric-magnetic dual formulation is completed by the relation
\be
\vec{B}=-\d_t \vec{C}.
\ee

{\bf Conserved ``helicities" in sourceless electromagnetism}

Sourceless electromagnetism has "helicities", quasi-topological quantities defined as 
spatial Chern-Simons forms, integrals over space of scalars made up from $\vec{A}$ and $\vec{C}$.
They are
\bea
H_{ee}&=&\int d^3x \vec{C}\cdot \vec{E}=\int d^3x \vec{C}\cdot \vec{\nabla}\times \vec{C}=\int d^3x \epsilon^{ijk}C_i\d_j C_k\cr
H_{mm}&=&\int d^3x \vec{A}\cdot \vec{B}=\int d^3x\epsilon^{ijk}A_i\d_j A_k\cr
H_{em}&=&\int d^3x \vec{C}\cdot\vec{B}=\int d^3x \epsilon^{ijk}C_i\d_j A_k\cr
H_{me}&=&\int d^3x \vec{A}\cdot\vec{E}=\int d^3x \epsilon^{ijk}A_i\d_k C_k\;,
\eea
where $H_{ee}$ is the electric helicity (Chern-Simons form of $\vec{C}$), $H_{mm}$ is its 
and its electromagnetic dual, the magnetic helicity (Chern-Simons form of $\vec{A}$),
$H_{em}$ is the electromagnetic helicity (BF form of $\vec{C}$ and $\vec{A}$) and 
$H_{me}$ is its electromagnetic dual, the magnetoelectric one. 

Note that these quantities are not guaranteed to be invariant in time, and correspondingly the result for them is not necessarily integer
either.

The variation in time of these helicities is
\bea
\d_t H_{mm}&=&\int d^3x (\d_t \vec{A}\cdot \vec{B}+\vec{A}\cdot \d_t\vec{B})=-2\int d^3x \vec{E}\cdot\vec{B}\cr
\d_t H_{ee}&=&\int d^3x (\d_t \vec{C}\cdot \vec{E}+\vec{C}\cdot \d_t\vec{E})=-2\int d^3x \vec{E}\cdot \vec{B}\cr
\d_t H_{me}&=& \int d^3x(\d_t \vec{A}\cdot \vec{E}+\vec{A}\cdot\d_t\vec{E})=-\int d^3x(\vec{E}^2-\vec{B}^2)\;,\cr
\d_t H_{em}&=& \int d^3x(\d_t \vec{C}\cdot \vec{B}+\vec{C}\cdot \d_t \vec{B})=\int d^3x (\vec{E^2}-\vec{B}^2).
\eea
Here we have used Maxwell's equations and partial integration. 
We see that if $\vec{E}\cdot \vec{B}=0$, $H_{mm}$ and $H_{ee}$ are conserved in time, whereas if $\vec{E}^2=\vec{B}^2$, 
$H_{me}$ and $H_{em}$ are conserved.

The case we are interested in this article is when both invariants vanish,
 $\vec{E}\cdot \vec{B}=0$ and also 
$\vec{E}^2-\vec{B}^2=0$, so all helicities are conserved.

In this case, there are "knotted" solutions, where $\vec{E}$ and $\vec{B}$ have a nonzero linking number. 
 
\subsection{Bateman's construction and knotted solutions}

These solutions are simplest in a construction due to Bateman. 

We can introduce the complex Riemann-Silberstein vector
\be
\vec{F}\equiv \vec{E}+i\vec{B}\;,
\ee
and in terms of it, 
the Maxwell's equations become
\be
\vec{\nabla}\times \vec{F}=i\frac{\d}{\d t}\vec{F};\;\;\; \vec{\nabla}\cdot \vec{F}=0.
\ee

To automatically satisfy the second equation, $\vec{\nabla}\cdot \vec{F}=0$, Bateman introduced an ansatz for $\vec{F}$,
\be
\vec{F}=\vec{\nabla}\a \times \vec{\nabla}\b.
\ee

It remains to solve the equation
\be\label{Batequ}
i\vec{\nabla}\times (\d_t \a \vec{\nabla}\b-\d_t\b\vec{\nabla}\a)=\vec{\nabla}\times \vec{F}\;,
\ee
where we have replaced the ansatz for $\vec{F}$  inside the time derivative, and taken outside the $\vec{\nabla}\times$. 
As we can see, we can solve the above equation if we peel off $\vec{\nabla}\times$, to obtain 
\be\label{Batequsol}
i (\d_t \a \vec{\nabla}\b-\d_t\b\vec{\nabla}\a)=\vec{F}=\vec{\nabla}\a\times \vec{\nabla}\b.
\ee
This is the only equation that needs to be satisfied in Bateman's construction. 

Expressing one $\vec{F}$ in one way, and another in the other, we find that $\vec{F}^2=0$, which means that
\bea
&&\vec{F}^2=i (\d_t \a \vec{\nabla}\b-\d_t\b\vec{\nabla}\a)(\vec{\nabla}\a\times \vec{\nabla}\b)=0\Rightarrow\cr
&&\vec{E}^2-\vec{B}^2=0,\;\;\;\; \vec{E}\cdot\vec{B}=0.
\eea
Now electromagnetic duality is just $\a\rightarrow i\a$ or $\b\rightarrow i\b$. 
For the solutions in Bateman's construction, since $\vec{F}^2=\vec{E}^2-\vec{B}^2+2i\vec{E}\cdot \vec{B}=0$,
the helicities $H_{ee}, H_{mm}, H_{em}$ and $H_{me}$ are all conserved. 

{\bf The Hopfion and knotted solutions for electromagnetism}

There is a simple solution of sourceless electromagnetism called a "Hopfion", for which the electric and magnetic fields are linked.  
In Bateman's construction, the Hopfion is defined by 
\bea
\a&=&\frac{A-1+iz}{A+it};\cr
\b&=&\frac{x-iy}{A+it}\cr
A&=& \frac{1}{2}(x^2+y^2+z^2-t^2+1).
\eea
We can check that it satisfies (\ref{Batequsol}). Moreover, replacing $\a\rightarrow \a^p$ and $\b\rightarrow \b^q$, we find 
more general $(p,q)$ knotted solutions. 

As we said, above, for any Bateman solution, the helicities are automatically conserved. But they need not be nonzero; in fact, one finds that 
for the Hopfion,
\be
H_{ee}=H_{mm}\neq 0;\;\;\; H_{em}=H_{me}=0. 
\ee

\section{Relativistic action and Hamiltonian formalism for Bateman's construction}

Maxwell's equations are of course linear. But the null condition $\vec{F}^2=0$, or $\vec{E}^2-\vec{B}^2=0$, $\vec{E}\cdot \vec{B}=0$
of Bateman's construction introduces a nonlinearity, that allows for the knotted null solutions to be nontrivial. Note that the sum of two 
null solutions is not generically null. If we have an electromagnetic wave propagating in the $z$ direction, this is a null solution. Adding two 
such null solutions, waves propagating {\em in the same direction} is a special case in which the sum of the null solutions is also null. 
Thus the null subspace is nontrivial, and propagates some degrees of freedom. It is therefore of interest to construct an action for 
this subspace. 

In  \cite{Hoyos:2015bxa}, a covariant formulation of the Bateman's construction was presented, which we quickly review. The covariant 
form of the Bateman ansatz is 
\be
F^{\mu\nu}-\frac{i}{2}\epsilon^{\mu\nu\rho\sigma}F_{\rho\sigma}=-\epsilon^{\mu\nu\rho\sigma}\d_\rho \a\d_\sigma\b.\label{covbate}
\ee
Indeed then, with the convention $\epsilon^{ijk0}=+\epsilon^{ijk}$, we obtain for the two components
\bea
E^i+iB^i&=& F^{0i}+\frac{i}{2}\epsilon^{ijk}F_{jk}=+\epsilon^{ijk}\d_j \a \d_k\b\cr
-i\epsilon^{ijk}(E^k+iB^k)&=&F^{jk}-i \epsilon^{jki}F^{0i}=-\epsilon^{jki}(\d_i\a \d_0 \b-\d_0\a\d_i\b)\;,
\eea
which means we get the two equivalent (via Maxwell's equations) forms of the non-covariant Bateman's ansatz,
\be
E^i+iB^i=\epsilon^{ijk}\d_j\a\d_k \b=i(\d_0\a\d_i\b-\d_i\a\d_0\b)\;,
\ee
which means that 
\bea
E^i&=& \epsilon^{ijk}{\rm Re}(\d_j\a \d_k\b)=-{\rm Im}(\d_0 \a\d_i\b-\d_i\a\d_0\b)\cr
B^i&=&\epsilon^{ijk}{\rm Im}(\d_j\a\d_k\b)={\rm Re}(\d_0\a\d_i\b-\d_i\a\d_0\b).\label{ebab}
\eea

Moreover, one can obtain the form of the potentials $\vec{A}$ and $\vec{C}$, or even covariantly, as
\bea
A_\mu&=&\frac{1}{2}{\rm Im}(\a\d_\mu \b-\b\d_\mu\a)\cr
C_\mu&=&\frac{1}{2}{\rm Re}(\a\d_\mu\b-\b\d_\mu\a).
\eea

We would like to construct an action that has (\ref{covbate}) among its equations of motion. Since $\a$ and $\b$ are complex, 
we cannot obtain a good action with just $F_{\mu\nu}$ added as an indepedent variable. We could consider $F_{\mu\nu}$ complex, and 
impose the reality condition with a Lagrange multiplier. Then the action will be
\bea
S&=&c\int d^4x \left[\frac{1}{2}\left(F^{\mu\nu}-\frac{i}{2}\epsilon^{\mu\nu\rho\sigma}F_{\rho\sigma}\right)^2+
\left(F_{\mu\nu}-\frac{i}{2}\epsilon_{\mu\nu\a\b}F^{\a\b}\right)\epsilon^{\mu\nu\rho\sigma} \d_\rho \a \d_\sigma\b\right.\cr
&&\left. +c.c.+\phi^{\mu\nu,\rho\sigma}(F_{\mu\nu}-\bar F_{\mu\nu})(F_{\rho\sigma}-\bar F_{\rho\sigma})
\right]\;,
\eea
where $c$ is a constant that we will fix later, and $\phi^{\mu\nu,\a\b}$ is a field with two independent sets of antisymmetric indices. 
We can check that if $F_{\mu\nu}$ is real, the action above reduces to (after writing explicitly the 
complex conjugate term)
\be
S=4c\int d^4x\left[\frac{1}{2}F_{\mu\nu}F^{\mu\nu}+\frac{1}{2}\epsilon^{\mu\nu\rho\sigma}F_{\mu\nu}{\rm Re}(\d_\rho \a\d_\sigma\b)
-F^{\rho\sigma}{\rm Im}(\d_\rho\a\d_\sigma\b)\right]\;,
\ee
which gives the equation of motion 
\be
F^{\mu\nu}={\rm Im}(\d^{[\mu}\a\d^{\nu]}\b)+\frac{1}{2}\epsilon^{\mu\nu\rho\sigma}{\rm Re}(\d_\rho\a\d_\sigma\b)\;,
\ee
which has also more general solutions than the ones of Bateman's construction.

If $F_{\mu\nu}$ is complex however, the $F_{\a\b}$ and $\bar F_{\a\b}$ equations of motion are 
\bea
&&\left(\delta_{\mu\nu}^{\a\b}-\frac{i}{2}{\epsilon_{\mu\nu}}^{\a\b}\right)
\left[F^{\mu\nu}-\frac{i}{2}\epsilon^{\mu\nu\rho\sigma}F_{\rho\sigma}+\epsilon^{\mu\nu\rho\sigma}\d_\rho \a\d_\sigma\b\right]\cr
&&+(\phi^{\mu\nu,\a\b}+\phi^{\a\b,\mu\nu})(F_{\mu\nu}-\bar F_{\mu\nu})=0\;,
\eea
and its complex conjugate. The $\phi^{\mu\nu,\a\b}$ equation, $(F_{\mu\nu}-\bar F_{\mu\nu})(F_{\a\b}-\bar F_{\a\b})=0$, 
then restricts to real $F_{\mu\nu}$. 
The $\a$ and $\b$ equations are 
\bea
&&\epsilon^{\mu\nu\rho\sigma}\d_\sigma\b\d_\rho\left(F_{\mu\nu}-\frac{i}{2}\epsilon_{\mu\nu\a\b}F^{\a\b}\right)=0\cr
&&\epsilon^{\mu\nu\rho\sigma}\d_\rho\a\d_\sigma\left(F_{\mu\nu}-\frac{i}{2}\epsilon_{\mu\nu\a\b}F^{\a\b}\right)=0\;,
\eea
and their complex conjugates. This just amounts to the Maxwell equations, $\d_\mu F^{\mu\nu}=0$ and $\d_{[\mu}F_{\nu\rho]}=0$, which 
also mean that we can write $F=dA$ and $*F=dC$, which was not imposed in the action. 

The essential feature above was the existence of two independent degrees of freedom of $F_{\mu\nu}$ time (the real and imaginary parts). 
But it was not essential that they be combined into a complex field. In fact, we will instead work with an  equivalent action, with two real fields, 
$F_{\mu\nu}$ and $\tilde F_{\mu\nu}$, imposing their equality by a Lagrange multiplier as before. 

The action is 
\bea
S&=&c\int d^4x \left[\frac{1}{2}F_{\mu\nu}\tilde F^{\mu\nu}-F^{\mu\nu}{\rm Im}(\d_{[\mu} \a\d_{\nu]}\b)+\frac{1}{2}\epsilon^{\mu\nu\rho\sigma}\tilde F_{\mu\nu}
{\rm Re}(\d_\rho\a\d_\sigma\b)\right.\cr
&&\left.-\phi^{\a\b,\mu\nu }(F_{\a\b}-\tilde F_{\a\b})(F_{\mu\nu}-\tilde F_{\mu\nu})\right].\label{batemanaction}
\eea
The equations of motion for $F_{\mu\nu}$ and $\tilde F_{\mu\nu}$ are 
\bea
&&\frac{1}{2} \tilde F_{\mu\nu}-{\rm Im} (\d_{[\mu}\a\d_{\nu]}\b)
-(\phi^{\a\b,\mu\nu}+\phi^{\mu\nu,\a\b})(F_{\a\b}-\tilde F_{\a\b})=0\cr
&& \frac{1}{2}F_{\mu\nu}+\frac{1}{2}{\epsilon_{\mu\nu}}^{\rho\sigma}{\rm Re}(\d_\rho\a\d_\sigma\b)
+(\phi^{\a\b,\mu\nu}+\phi^{\mu\nu,\a\b})(F_{\a\b}-\tilde F_{\a\b})=0\;,\cr
&&
\eea
and the $\phi^{\mu\nu, \a\b}$ equation of motion is $(F_{\mu\nu}-\tilde F_{\mu\nu})(F_{\a\b}-\tilde F_{\a\b})=0$, 
killing the last term, and leaving only the desired equations.

The equations of motion for $\a,\bar\a,\b,\bar\b$ are 
\bea\label{EOM}
&&\d_\sigma\b\left[-i\d_\rho F^{\rho\sigma}-\frac{1}{2}\epsilon^{\mu\nu\rho\sigma}\d_\rho\tilde F_{\mu\nu}\right]=0\cr
&&\d_\sigma\a\left[-i\d_\rho F^{\rho\sigma}-\frac{1}{2}\epsilon^{\mu\nu\rho\sigma}\d_\rho\tilde F_{\mu\nu}\right]=0
\eea
and their complex conjugates, which again gives just the Maxwell's equations $\d_\mu F^{\mu\nu}=0$ and $\d_{[\mu}F_{\nu\rho]}=0$.

Next we would like to check what is the value of the Lagrangain density on shell, namely when one substitute into it the various fields that are solutions of the equations of motion. 
It is easy to realize that 
\be
{\cal L}_{on\ shell} = -\frac12 F_{\mu\nu}F^{\mu\nu} =0
\ee
In particular it vanishes due to the equations of motion \ref{EOM}. This is of course in full accordance with the null solutions for which 
\be
F_{\mu\nu}F^{\mu\nu}= \frac12 (\vec E^2 - \vec B^2)= 0 \qquad \tilde F_{\mu\nu}F^{\mu\nu}= \vec E\cdot\vec B=0
\ee
To write the Hamiltonian, we first calculate the canonically conjugate momenta, obtaining
\bea
p_\a&=&\frac{c}{2}\left[iF^{0i}+\frac{1}{2}\epsilon^{jk0i}\tilde F_{jk}\right]\d_i\b\cr
p_{\bar \a}&=&\frac{c}{2}\left[-iF^{0i}+\frac{1}{2}\epsilon^{jk0i}\tilde F_{jk}\right]\d_i\bar\b\cr
p_\b&=&\frac{c}{2}\left[iF^{i0}+\frac{1}{2}\epsilon^{jki0}\tilde F_{jk}\right]\d_i\a\cr
p_{\bar \b}&=&\frac{c}{2}\left[-iF^{i0}+\frac{1}{2}\epsilon^{jki0}\tilde F_{jk}\right]\d_i\bar\a\cr
p_{F^{0i}}&=&c{\rm Im}(\a\d_i\b-\b\d_i\a)\cr
p_{\tilde F^{ij}}&=&\frac{c}{2}\epsilon^{ij0k}{\rm Re}(\a\d_k\b-\b\d_k\a)\;,
\eea
and the rest are zero.
Note then that all the above momenta are actually (primary) constraints, since they don't involve time derivatives. Moreover, we also have the 
extra constraints
\be
p_{F^{ij}}=p_{\tilde F^{0i}}=p_{\phi^{\mu\nu,\a\b}}=0.
\ee

Then the classical Hamiltonian is 
\bea
H&=&\int d^3x [p_\a \dot \a+p_\b \dot \b +p_{\bar\a}\dot{\bar\a}+p_{\bar\b}\dot{\bar\b}+p_{F^{0i}}\dot F^{0i}+
p_{\tilde F^{ij}}\dot{\tilde F}^{ij}-{\cal L}]\cr
&=& c\int d^3x \left[F^{0i}\tilde F^{0i}-\frac{1}{2}F^{ij}\tilde F^{ij}+\frac{1}{c}\dot F^{0i}p_{F^{0i}}+\frac{1}{c}\dot{\tilde F}^{ij}p_{\tilde F^{ij}}\right.\cr
&&\left.+F^{ij}{\rm Im}(\d_i\a\d_j\b)+\epsilon^{ijk}\tilde F_{k0}{\rm Re}(\d_i\a\d_j\b)+\phi^{\a\b,\mu\nu}(F_{\a\b}-\tilde F_{\a\b})
(F_{\mu\nu}-\tilde F_{\mu\nu})\right].
\eea
On the $\phi$ equation of motion, replacing the momenta with their expression in terms of fields, and doing a partial integration, we obtain
\bea
H&=&c\int d^3x \left[(F^{0i})^2-\frac{1}{2}(F^{ij})^2-F^{0i}{\rm Im}(\d_0\a\d_i\b-\d_i\a\d_0\b)+F^{ij}{\rm Im}(\d_i\a\d_j\b)\right.\cr
&&\left.+\epsilon^{ijk}F^{0k}{\rm Re}(\d_i\a \d_j\b)+\frac{1}{2}\epsilon^{ijk}F^{ij}{\rm Re}(\d_0\a\d_k\b -\d_0\b\d_k\a)\right].
\eea
On shell, replacing (\ref{ebab}) in the above, we obtain 
\be
H=c\int d^3x (3\vec{E}^2+\vec{B}^2).
\ee
Since the Hamiltonian density on-shell is supposed to be $(\vec{E}^2+\vec{B}^2)/2$, but we have $\vec{E}^2-\vec{B}^2=0$ on-shell,
we find that $c=1/4$.

\section{Symmetries and conserved charges}

In this section, we will investigate the symmetries of the action, having in mind a possible generalization of the "null subsector"
described by our action to other systems. We then construct such null subsectors of other theories.

\subsection{Symmetries  of the action for Bateman's construction  and conserved charges}

The action for the Bateman formulation has the following symmetries and charges:
\begin{itemize}
\item
By construction, the action is invariant under the Poincar\'{e} group. Moreover, since it does not include any scale, 
it is actually invariant under the full $SO(2,4)$ conformal group.
\item
The action includes only derivatives of the complex scaler fields, and therefore it is invariant under
\be\label{shiftsym}
\a(x^\mu) \rightarrow \a(x^\mu)+ a \qquad \b(x^\mu) \rightarrow \b(x^\mu)+ b
\ee
where $ a ,b$ are constant complex numbers.
\item
In addition, the theory is characterized by a set of four helicities, which are conserved without being 
affiliated with symmetry transformations of the action.
\end{itemize}

\subsection{The energy-momentum tensor and Noether charges associated with the conformal symmetry}

The Noether currents associated with the full conformal group are, as is well known, all built from the energy-momentum tensor. 
We can determine the latter either by using the standard Noether procedure (obtaining the Noether energy-momentum tensor, 
and then symmetrizing), or by coupling the system to an external metric and varying the action with respect to it (obtaining the 
Belinfante tensor). In both ways, the result is the same, and the energy-momentum tensor of the system is found to be:
\bea
T_{\mu\nu} &=& -{F_\mu}^\lambda{\rm Im}(\d_{[\lambda} \a\d_{\nu]}\b)+\frac{1}{2}\tilde F_{\mu\lambda}{\epsilon_\nu}^{\rho\sigma\lambda} 
{\rm Re}(\d_\rho\a\d_\sigma\b)-\eta_{\mu\nu} {\cal L}\cr
&=& {\epsilon_\mu}^{\rho\sigma\lambda}  {\rm Im}(\d_{[\lambda} \a\d_{\nu]}\b)  
{\rm Re}(\d_\rho\a\d_\sigma\b) 
\eea
The last expression follows from the fact that  on shell  ${\cal L}=0$. One can then easily check that we have
\be
{T^\mu}_\mu=0\;,\;\;\;
{T^\mu}_\lambda {T^\lambda}_\nu=0\;,
\ee
and in fact for every $n\geq 2$,
\be
{(T^n)^\mu}_\nu=0\;,
\ee
in matrix sense. 

We can take this to be a more general condition that the null condition $\vec{F}^2=0$, one that is both covariant 
(the null condition can be expressed covariantly also as $F_{\mu\nu}F^{\mu\nu}=\epsilon^{\mu\nu\rho\sigma}F_{\mu\nu}F_{\rho\sigma}=0$),
and general enough so that it can be applied to other cases. 

In particular,  \cite{Alves:2017ggb} embedded the same null Hopfion solution in fluid dynamics, with $P=0$, for which the 
energy-momentum tensor is 
\be
T_{\mu\nu}=\rho u_\mu u_\nu\;,
\ee
and the velocity is null, $u^\mu u_\mu=0$. As we see, in this case also we have the same condition 
\be
{T^\mu}_\mu=0\;,\;\;\; {(T^n)^\mu}_\nu=0\;.\;\;\; \forall n\geq 2.
\ee
This suggests that we can generalize the condition to other systems. However, it could be that in some cases 
we need to impose the weaker condition 
\be
\Tr[{(T^n)^\mu}_\nu]=0.
\ee

It is also easy to check that the vanishing of the quadratic form built from the energy-momentum tensor guarantees 
also that the same is true for the whole set of Noether currents associated with the conformal group,
\be
J^\mu_{\alpha\beta} J_{\mu\gamma\delta}= D_\mu D^\mu = K_{\mu\lambda} K^\lambda_\nu=0\;,
\ee
where $J^\mu_{\alpha\beta}, D_\mu, K_{\mu\nu}$ are the currents associated with Lorentz transformations, scale 
transformations and special conformal transformations, respectively.


\section{Other null systems}

We can now impose the vanishing of ${(T^n)^\mu}_\nu$ as a condition for other systems, and construct the Lagrangean for their 
null subsector.

\subsection{Null free massless complex scalar}
Consider the case of a free massless scalar, 
\be
{\cal L}=-(\d_\mu\phi) (\d^\mu\phi^*).
\ee
The (Belinfante) energy-momentum tensor is 
\be
{T^\mu}_\nu=(\d^\mu\phi \d_\nu \phi^* +\d^\mu\phi^* \d_\nu \phi)  -\delta^\mu_\nu (\d^\mu\phi\d_\mu\phi^*).
\ee
In analogy with the case of electromagnetism, we define the null configuration by the vanishing of the Lagrangean,
\be
{\cal L}=-(\d_\mu\phi) (\d_\mu\phi^*) =0.
\ee
We can check that this is the only way in which we can have 
\be 
{T^\mu}_\mu= 0
\ee 
{\em off-shell}. For these configurations, we have  more generally
\be
T_{\mu\nu} T^{\nu\rho}=0\;, \qquad T_{\mu_1\mu_2} T^{\mu_2\mu_3}....T_{\mu_{n-1}\mu_{n}} T^{\mu_n,\mu_{n+1}} =0\;,\;\;\forall n,
\ee 

It is easy to find simple solutions of the null condition above. In fact, for a null $k^\mu$, $k^2=0$, any function of $k\cdot x$ will be null and 
on-shell: $\d_\mu^2\phi=0$ and $\d_\mu \phi \d^\mu\phi^*=0$. Natural examples are:
\be
(k\cdot x)^n\;,\;\;\; \forall n\;,\;\;\; {\rm and}\;\;\;\; e^{ik\cdot x}.
\ee

Inspired by the electromagnetic case, where this was valid and was used to find new solutions in \cite{Hoyos:2015bxa}, 
we check whether {\em complex} conformal transformations is a symmetry on the space of null solutions, so helps us generate new ones. 

Consider an infinitesimal {\em complex} coordinate transformation, $\delta x^\mu =\xi^\mu$, under which the complex scalar $\phi$
transforms as 
\be
\delta \phi=\xi^\lambda \d_\lambda \phi.
\ee
For a {\em complex conformal} transformation, $\phi$ gets an additional transformation, since a 
scalar field transforms under conformal transformations as
\be
\phi(x)\rightarrow \phi(x')\left|\frac{\d x'^\mu}{\d x^\nu}\right|^{-\Delta/d}\;,
\ee
where $\Delta$ is the conformal dimension, and $\Delta =1$ for the massless scalar in 3+1 dimensions. Then we 
have\footnote{The determinant becomes 
infinitesimally $\simeq 1+\d_\mu \xi^\mu$, leading to the written formula.}
\be
\delta_{\rm conf.}\phi=\xi^\lambda \d_\lambda \phi-\frac{\Delta}{d}\phi \d_\lambda \xi^\lambda.
\ee

Transforming the equation of motion $\d^2\phi=0$, we obtain a term with $\xi^\lambda\d_\lambda$ on the equation, vanishing on-shell, and 
the vanishing of the other terms gives the condition 
\bea
&&2(\d_\mu \xi^\lambda)(\d^\mu\d_\lambda\phi)+(\d_\lambda\phi)\d_\mu\d^\mu\xi^\lambda\cr
&&-\frac{\Delta}{d}[2(\d_\lambda\phi)\d^\lambda(\d_\mu \xi^\mu)+\phi\d^2(\d_\mu\xi^\mu)]=0.
\eea

The conformal group in 4 dimensions contains translations, Lorentz rotations, scalings and special conformal transformations. 

-For translations $\xi^\lambda=c^\lambda$, the condition above is obviously satisfied. 

-For Lorentz rotations $\xi^\lambda={\Lambda^\lambda}_\rho x^\rho$, we obtain 
$2{\Lambda^\lambda}_\rho\d^\rho \d_\lambda \phi=0$, which is satisfied since ${\Lambda^\lambda}_\rho=-{\Lambda^\rho}_\lambda$.

-For scalings, $\xi^\lambda =\lambda x^\lambda$, we obtain $2\lambda\d^2\phi=0$, satisfied on-shell.

-For special conformal transformations, 
\be
\xi^\lambda=a^\lambda x^2-2a_\rho x^\rho x^\lambda\;,
\ee
after a bit of algebra, the condition becomes
\be
4\left(1-\frac{4\Delta}{d}\right)a^\lambda \d_\lambda \phi=0\;,
\ee
satisfied for $\Delta=1$ and $d=4$.

We next check whether the complex conformal group leaves invariant the null condition $(\d_\mu\phi)(\d^\mu\phi^*)=0$. 
Varying the condition, and dropping terms proportional to the (derivative of the) condition itself, we obtain the condition
\bea
&&(\d_\mu\phi^*)(\d^\mu\xi^\lambda)(\d_\lambda\phi)\cr
&&-\frac{\Delta}{d}(\d_\mu\phi^*)\phi \d_\mu(\d_\lambda\xi^\lambda)+c.c=0.
\eea

-For translations it is obviously satisfied.

-For Lorentz rotations, we obtain the condition 
\be
(\d^\mu\phi^*)(\d_\lambda\phi) {\Lambda^\lambda}_\mu+c.c.=0\;,
\ee
which is satisfied for ${\Lambda^\lambda}_\rho=-{\Lambda^\rho}_\lambda$.

-For scalings, the condition becomes $\lambda (\d_\mu\phi^*)(\d^\mu\phi)=0$, satisfied.

-For special conformal transformations, after a bit of algebra, assuming the null condition itself, the condition becomes
\be
\frac{8\Delta}{d}\phi(a^\mu\d_\mu\phi^*)+c.c.=0.
\ee

That means that only special conformal transformations with parameter $a^\mu$ satisfying $a^\mu \d_\mu\phi=0$ leave invariant the 
null condition. 

We can then use the complex conformal transformations satisfying the above constraint to find new solutions. 
Note that the inversion, used in the electromagnetism case, does not satify the condition: $\delta x^\mu=\xi^\mu\propto x^\mu$. 

Indeed, for instance the power law solution, $(k\cdot x)^n$, with $k^2=0$, becomes after the inversion (up to an irrelevant constant),
\be
\phi=\frac{(-k_0t+\vec{k}\cdot \vec{x})^n}{(r^2-t^2)^{n-1}}.
\ee
We can then check explicitly that $\d_\mu\phi\d^\mu\phi^*=0$ if and only if $n=1$, in which case however the solution is unchanged by the 
inversion.

On the other hand, we would have wanted $\phi\rightarrow 0$ as $r\rightarrow \infty$, true only if  if $n\geq 3$, which 
means that in this case, we can identify the points at infinity, and effectively 
compactify $\mathbb{R}^3$ space to $S^3$. Moreover, then at $t=0$, $\phi$ is never infinite, effectively compactifying also the complex plane 
$\phi\in \mathbb{C}$ to $S^2$. Thus $\phi$ becomes a map $\phi:S^3\rightarrow S^2$, which is characterized by a Hopf index.

But we can consider the same solution, just translated in time, $t\rightarrow t-i$, and divided by the norm squared, i.e.,
\be
\phi=\frac{(-k_0(t-i)+\vec{k}\cdot\vec{x})^n}{|(-k_0(t-i)+\vec{k}\cdot\vec{x})^n|^2}.
\ee
This still has the properties above, so defines a map $\phi:S^3\rightarrow S^2$, characterized by a Hopf index.

The topological charge associated with solutions equivalent to maps $\phi: S^3\rightarrow S^2$ is the Hopf charge. For a complex 
scalar $\phi$, consider the "field strength"
\be
F_{ij}=\d_i A_j -\d_j A_i =\frac{1}{4\pi i}\frac{\d_i \phi\d_j\phi^*-\d_i\phi^*\d_j \phi}{(1+|\phi|^2)^2}\;,
\ee
coming from the "gauge field"
\be
A_i=\frac{-1}{8\pi i}\frac{\d_i \ln \phi-\d_i\ln \phi^*}{1+|\phi|^2}.
\ee
Then the Hopf index (charge) is (see \cite{Alves:2017zjt} for more details)
\be
N=\int_{S^3}F\wedge A=
\int_{S^3}\frac{d^3x}{32\pi^2}\epsilon^{ijk}\frac{(\d_i\ln \phi-\d_i\ln\phi^*)(\d_i\phi\d_j\phi^*-\d_i\phi^*\d_j\phi)}{(1+|\phi|)^3}.
\ee

A related quantity is obtained by projecting $\phi$ (defined on $\mathbb{C}^*$) onto a vector $\vec{n}$ on $S^3$ in Euclidean coordinates. 
The map is done using the standard stereographic projection
\be
n_1+in_2=\frac{2\phi}{1+|\phi|^2}\;,\;\;\; n_3=\frac{-1+|\phi|^2}{1+|\phi|^2}=1-\frac{2}{1+|\phi|^2}\;,
\ee
where $n_i$ are the Euclidean coordinates of $\vec{n}$ (with $\vec{n}^2=1$). 

Then the topological quantity related to the Hopf charge is 
\be
N'=\frac{1}{24\pi^2}\int_{S^3} \epsilon^{ijk}\epsilon_{abc}\d_i n^a \d_j n^b \d_k n^c.
\ee

The quantities $N$ and $N'$ play the role of the helicities and Hopf charge in the electromagnetic case.

Note that the null condition we have defined in this section is related to the one for the fluid, since for a (real) scalar, we can define a (fluid) 4-velocity 
(see \cite{Nastase:2015ljb} for a more detailed analysis)
\be
u^\mu\propto \d^\mu \phi\;,
\ee
so $|\d_\mu\phi|^2=0$ means $u_\mu u^\mu=0$, the same condition as for the fluid. 

We can then also easily construct a Lagrangean for the null subsector: we just need to add a Lagrange multiplier $\lambda$ to the coefficient of the 
action, 
\be
{\cal L}_{\rm null}=-(1+\lambda)|\d_\mu\phi|^2.
\ee

\subsection{ Null dual 2-form field}

We can consider now the (4 dimensional) 
Poincar\'{e} dual of the free massless scalar, a 2-form gauge field $B_{\mu\nu}$ with 3-form field strength
\be
H_{\mu\nu\rho}=3\d_{[\mu}B_{\nu\rho]}\;,
\ee
where the antisymmetrization is with strength one. The Lagrangean for the gauge field is 
\be
{\cal L}=-\frac{|H_{\mu\nu\rho}|^2}{3!}\;,
\ee
and the field is Poincar\'{e} dual to the scalar above, via
\be
H_{\mu\nu\rho}=i\epsilon_{\mu\nu\rho\sigma}\d^\sigma \phi.
\ee
Replacing back in the action, we obtain the scalar action. 

Now the energy-momentum tensor is 
\be
{T^\mu}_\nu=(H^{\mu\rho\sigma}H_{\nu\rho\sigma}^*+H^{*\mu\rho\sigma}H_{\nu\rho\sigma})-\delta^\mu_\nu|H_{\rho\sigma\lambda}|^2.
\ee
The condition ${T^\mu}_\mu=0$ again implies
\be
|H_{\mu\nu\rho}|^2=0\;,
\ee
which is the same condition as for the scalar, $|\d_\mu\phi|^2=0$. Replacing back in the energy-momentum tensor, we have 
\be
{T^\mu}_\nu=(H^{\mu\rho\sigma}H_{\nu\rho\sigma}^*+H^{*\mu\rho\sigma}H_{\nu\rho\sigma})\;,
\ee
and it would seem that we would need to impose the weaker condition on $\Tr[T^n]$ instead of $T^n$, to have it automatically satisfied, 
but in fact, by changing variables to $\phi$ in ${T^\mu}_\nu$, we see that $|H_{\mu\nu\rho}|^2=-3!|\d_\mu\phi|^2=0$ is enough to 
satisfy $T^n=0$. 

Thus the null subsector can be defined also in the same way as for the scalar, by 
adding a Lagrangian multiplier $\lambda$ to the coefficient of the action,
\be
{\cal L}=-(1+\lambda)\frac{|H_{\mu\nu\rho}|^2}{3!}.
\ee

\subsection{ Null massless DBI scalar}

Consider now a massless DBI scalar (a scalar version of Born-Infeld electromagnetism \cite{Born:1934gh}, which 
appears as the action for a fluctuating D-brane in string theory, and was used for instance in Heisenberg's model \cite{Heisenberg:1952zz}
for saturation of Froissart's unitarity bound \cite{Froissart:1961ux} as an action for the pion in this asymptotic limit; see \cite{Nastase:2015ixa}
for generalizations), with Lagrangian 
\be
{\cal L}=l^{-4}\left[1-\sqrt{1+l^4|\d_\mu\phi|^2}\right].
\ee
The energy-momentum tensor is 
\be
{T^\mu}_\nu=\frac{\d^\mu\d_\nu\phi}{\sqrt{1+l^4(\d_\rho\phi)^2}}+\delta^\mu_\nu l^{-4}\left[1-\sqrt{1+l^4(\d_\mu\phi)^2}\right]+c.c..
\ee

In this case, we can also define the null condition like for the fluid, again taking advantage of the fact that 
$u^\mu\propto \d^\mu\phi$ is a fluid 4-velocity. Then the null condition $u^\mu u_\mu=0$ again becomes
$|\d_\mu\phi|^2=0$.

Indeed, the condition ${T^\mu}_\mu=0$ implies now also $|\d_\mu\phi|^2=0$. On the condition, we have 
\be
T_{\mu\nu}=\d_\mu \phi\d_\nu\phi^*+c.c.\;,
\ee
just like for the free massless scalar, so again we find that $({T^\mu}_\nu)^n=0$ is satisfied automatically. 

Now imposing the null condition with a Lagrange multiplier $\lambda$, we find the Lagrangean for the null sector,
\be
{\cal L}=l^{-4}\left[1-\sqrt{1+l^4|\d_\mu\phi|^2}\right]+\lambda|\d_\mu\phi|^2.
\ee

\subsection{Helicities}

Since in our new formulation, there is no more $\vec{E}$ and $\vec{B}$, we need to describe the topological 
quantities, i.e. helicities, in terms of the new variables of our action.

These non-Noether conserved charges, i.e. the helicities, can be expressed as
\be
H_{ab}= \int d^3x e^{ijk} P_a(\a\d_i\b-\b\d_i\a)P_b(\d_j\a\d_k\b-\d_k\b\d_j\a)
\ee
where  the indices $a,b$ stand for (magnetic, electric) and  $P_a$ are the projectors to imaginary and real parts $P_a= Im$ or $P_a= Re$.
We can check that these helicities are conserved by taking the time derivatives and making use of the equations of motion.

\section{Quantization of the system and observables}

To construct the quantum theory of the system, two obvious ways are the path integral formalism in the Lagrangian formulation, 
and the Dirac formalism in the Hamiltonian formulation.

\subsection{Path integral in Lagrangean formulation and observables}

Since we have a Lagrangian, and thus an action, for the Bateman's construction, we can certainly formally write a path integral. 
Since as we saw, the gauge field $A_\mu$ or its dual $C_\mu$ are not needed for the action, we integrate only over the 
fields whose equations of motion we wrote, namely $F_{\mu\nu}, \tilde F_{\mu\nu},\a,\b,\phi^{\mu\nu,\a\b}$, and obtain 
\be
Z=\int {\cal D}F_{\mu\nu}{\cal D}\tilde F_{\mu\nu}{\cal D}\a{\cal D}\b {\cal D}\phi^{\mu\nu,\a\b} e^{iS}\;,
\ee
where the action is given in (\ref{batemanaction}). But his partition function doesn't depend on anything, so it is not very useful; 
we must insert something inside the path integral. 

Defining as usual $E^i=F^{0i}$ and $B^i=\frac{1}{2}\epsilon^{ijk}F_{jk}$, we can construct the observable associated with two 
loops $C_1$ and $C_2$,
\be
\exp \left[\oint_{C_1}\vec{E}\cdot d\vec{x}\right]\exp\left[\oint_{C_2}\vec{B}\cdot d\vec{x}\right].
\ee
We know that, in the Hopfion type solution (and not in the propagating wave solutions), 
the electric and magnetic fields at fixed time are knotted for two loops with nonzero linking number. 

That means that, if we consider the observable 
\be
Z[t;C_1,C_2]\equiv \int {\cal D}F_{\mu\nu}{\cal D}\tilde F_{\mu\nu}{\cal D}\a{\cal D}\b {\cal D}\phi^{\mu\nu,\a\b} e^{iS}
\exp \left[\oint_{C_1}\vec{E}(t)\cdot d\vec{x}\right]\exp\left[\oint_{C_2}\vec{B}(t)\cdot d\vec{x}\right]\;,
\ee
there is at least a possibility for this to be nonzero if $C_1$ and $C_2$ are linked. On the other hand, if they are not linked, it 
seems improbable that the result is nonzero. 

We leave the calculation of these observables for further work, but we note the formal similarity 
with the famous case of Witten's Chern-Simons quantum field theory solution for the Jones polynomial \cite{Witten:1988hf}, where 
the polynomials are found from the path integral observable
\be
Z[M; C_i,R_i]\equiv \int {\cal D}A_\mu e^{iS} \prod_{i=1}^r W_{R_i}[C_i]\;,
\ee
and where $M$ is  a manifold, $C_i$ are loops, $R_i$ representations, $S$ is the Chern-Simons action, 
and $W_{R}[C]=\Tr_RP\exp [\oint_C A\cdot d x]$ is the Wilson loop.

\subsection{Dirac formalism in Hamiltonian formulation}

An alternative for constructing the quantum theory of this system is to construct the secondary constraints, 
calculate their Poisson brackets, and then construct Dirac brackets. 

The secondary constraints,  obtained from commuting the primary constraints with the Hamiltonian, are 
\bea
0=-\{p_{F^{ij}},\frac{1}{c}H\}&=&-\tilde F^{ij}+{\rm Im}(\d_{[i}\a\d_{j]}\b)+(\phi^{ij,\mu\nu}+\phi^{\mu\nu,ij})(F_{\mu\nu}-\tilde F_{\mu\nu})\cr
0=-\{p_{\tilde F^{0i}},\frac{1}{c}H\}&=&F^{0i}+\epsilon^{ijk}{\rm Re}(\d_i\a \d_j \b)-(\phi^{0i,\mu\nu}+\phi^{\mu\nu,0i})(F_{\mu\nu}-\tilde F_{\mu\nu})\cr
0=-\{p_{\phi^{\mu\nu,\rho\sigma}},\frac{1}{c}H\}&=&(F_{\mu\nu}-\tilde F_{\mu\nu})[F_{\rho\sigma}-\tilde F_{\rho\sigma}]\cr
0&=&-\{p_\a-\frac{c}{2}\left[iF^{0i}-\frac{1}{2}\epsilon^{ijk}\tilde F_{jk}\right]\d_i\b,\frac{1}{c}H\}\cr
&=&
\d_i\b\left[\frac{i}{2}(\d_i F^{ij}+\d_0 F^{0j})-\frac{1}{4}\epsilon^{ijk0}(2\d_i \tilde F^{0k}-\d_0 \tilde F^{ik})\right]\cr
0&=&-\{p_{\bar\a}-\frac{c}{2}\left[-iF^{0i}-\frac{1}{2}\epsilon^{ijk}\tilde F_{jk}\right]\d_i\bar\b,\frac{1}{c}H\}\cr
&=&
\d_i\bar\b\left[-\frac{i}{2}(\d_i F^{ij}+\d_0 F^{0j})-\frac{1}{4}\epsilon^{ijk0}(2\d_i \tilde F^{0k}-\d_0 \tilde F^{ik})\right]\cr
0&=&-\{p_\b+\frac{c}{2}[iF^{0i}-\frac{1}{2}\epsilon^{ijk}\tilde F_{jk}]\d_i\a,\frac{1}{c}H\}\cr
&=&
=-\d_i\a\left[\frac{i}{2}(\d_i F^{ij}+\d_0 F^{0j})-\frac{1}{4}\epsilon^{ijk0}(2\d_i \tilde F^{0k}-\d_0 \tilde F^{ik})\right]\cr
0&=&-\{p_{\bar\b}+\frac{c}{2}[-iF^{0i}-\frac{1}{2}\epsilon^{ijk}\tilde F_{jk}]\d_i\bar\a,\frac{1}{c}H\}\cr
&=&
=-\d_i\bar\a\left[-\frac{i}{2}(\d_i F^{ij}+\d_0 F^{0j})-\frac{1}{4}\epsilon^{ijk0}(2\d_i \tilde F^{0k}-\d_0 \tilde F^{ik})\right]\cr
0&=&-\{p_{F^{0i}}-c{\rm Im}(\a \d_i\b-\b\d_i\a),\frac{1}{c}H\}\cr
&=&
\tilde F^{0i}-\frac{1}{c}\frac{d}{dt}p_{F^{0i}}+(\phi^{0i,\mu\nu}+\phi^{\mu\nu,0i})(F_{\mu\nu}-\tilde F_{\mu\nu})\cr
0&=&-\{p_{\tilde F^{ij}}+\frac{c}{2}\epsilon^{ijk}{\rm Re}(\a\d_k\b-\b\d_k\a),\frac{1}{c}H\}\cr
&=&
-\frac{1}{2}F^{ij}-\frac{1}{c}\frac{d}{dt}p_{\tilde F^{ij}}-(\phi^{ij,\mu\nu}+\phi^{\mu\nu,ij})(F_{\mu\nu}-\tilde F_{\mu\nu}).
\eea
After some rearrangements, and using some of the constraints themselves on the others, these become just the full set of the equations 
of motion,
\bea
0&=&\tilde F^{\mu\nu}-{\rm Im}(\d^{[\mu}\a\d^{\nu]}\b)+(\phi^{\mu\nu,\rho\sigma}+\phi^{\rho\sigma,\mu\nu})(F_{\rho\sigma}-\tilde F_{\rho\sigma})\cr
0&=&F^{\mu\nu}-\epsilon^{\mu\nu\rho\sigma}{\rm Re}(\d_\rho\a\d\sigma\b)-(\phi^{\mu\nu,\rho\sigma}+\phi^{\rho\sigma,\mu\nu})(F_{\rho\sigma}-
\tilde F_{\rho\sigma})\cr
0&=&(F_{\mu\nu}-\tilde F_{\mu\nu})(F_{\rho\sigma}-\tilde F_{\rho\sigma})\cr
0&=&\d_\mu F^{\mu\nu}\cr
0&=&\d_{[\mu}\tilde F_{\nu\rho]}.
\eea

The primary constraints of the theory are the expressions for all the momenta as a function of the fields. We have then calculated that the secondary 
constraints contain all the equations of motion of the theory. 
Moreover, all the constraints have nontrivial Poisson brackets with some other constraint (they don't vanish weakly), 
since the primary constraints all involve the momenta, and the secondary constraints all involve the fields (resulting in terms with delta functions 
for their Poisson brackets), which means that 
all constraints are second class. The construction of the Dirac brackets however is very involved, and are left for further work.

\section{Observations on semiclassical quantization}

Bateman's formulation was very convenient in constructing the electromagnetic knot solutions, like the Hopfion and its generalizations 
\cite{Trautman:1977im,Ranada:1989wc,ranada1990knotted,Kedia:2013bw,besieris2009hopf,Hoyos:2015bxa}. 
These are solutions of the classical Maxwell's equations obeying constraints, which in the formulation of present paper 
become classical solutions of the equations of motion that have some topological properties. 

Semiclassical quantization of solitons, or more generally of special classical solutions, is usually done by introducing fluctuations around
these special classical solutions and quantizing these.  One can write an expansion on fluctuations with time dependent coefficients, and write
\be
\phi(x,t)=\phi_{\rm sol}(x)+\sum_{n\in\mathbb{N}}q_n(t)\eta_n(x)\;,
\ee
where $\eta_n(x)$ are eigenfunctions of the kinetic operator around the solution, less the time derivatives.
One then identifies the "collective coordinates" around the classical solutions, "$q_0$" or more generally $q^A$, which are 
zero energy modes, i.e., global symmetries of the solution (this method of "collective coordinate quantization was introduced in 
\cite{Christ:1975wt,Gervais:1974dc,Gervais:1975pa}, reviewed in \cite{Gervais:1975yg}). 
For instance, in most cases we have the position $X_0$, and introducing the mode amounts to a 
shift $x\rightarrow x-X_0$. Then one makes them time dependent, $X_0=X_0(t)$, which means that they cease to be symmetries, and have an energy 
that goes like a power of the velocity, $E\propto (\dot X^2)^n$. Moreover, now the solution with $x-X_0(t)$ is only a solution to the lowest order in 
$\dot X$, and one needs to correct the solution with extra terms in order to find a solution to higher orders. 
Expanding the (approximate) classical solution $\phi_{\rm sol}(x-X_0(t))$ in the perturbation $X_0(t)$, we find 
\be
\phi_{\rm sol}(x-X_0(t))\simeq \phi_{\rm sol}(x)-X_0(t)\phi'_{\rm sol}(x)+...
\ee
That means that we can make (for the linearized fluctuations) a change of basis, and remove $n=0$ from the sum over modes, and replace the $x$  
dependence with the $x-X_0(t)$ dependence, so 
\be
\phi(x,t)=\phi_{\rm sol}(x-X_0(t))+\sum_{m=1}^\infty q_m(t) \eta_m(x-X_0(t)).
\ee
By this form, we make a "change of basis" from the field $\phi(x,t)=\phi_x(t)$ to the infinite basis of quantum mechanical variables
$q^I(t)=(q_0=q^A=X_0(t),q_m(t))$. Substituting this expansion in the Hamiltonian of the system, written in quantum mechanics with the momentum
\be
\pi^x(t)=i\hbar\frac{\d}{\d\phi_x(t)}\;,
\ee
we find it in the form
\be
H=\frac{1}{2}g^{IJ}(q)p_I p_J+V(q)\;,
\ee
where  $p_I=i\hbar \d/\d q^I$, and at the classical level, the metric is 
\be
g_{IJ}=\int dx\frac{\d\phi(x,t)}{\d q^I}\frac{\d \phi(x,t)}{\d q_J}.
\ee
Since $X_0(t)$ appears only in the combination $x-X_0(t)$, the metric is independent of it, i.e. $g_{IJ}=g_{IJ}(q_m)$. But the Hamiltonian depends 
on $p_0$, and the metric for it is 
\be
g_{00}=\int dx [\phi'(x,t)]^2.
\ee
One can then proceed to quantize this quantum mechanical system (time dependent variables) 
based on the classical Hamiltonian and Poisson brackets, which is 
far easier than quantizing a full classical field. 

If we try to apply this formalism in our case, we note first that we have two obvious collective coordinates. Indeed, our Hopfion solutions in the
Bateman parametrization, as we saw, 
have a symmetry $\a\rightarrow \a+a$ and $\b\rightarrow \b+b$, so $a$ and $b$ are collective coordinates. We make them variables $a(t)$ and $b(t)$, 
together making up the $q_0=q^A=(a,b)$ of the general formalism. But then 
\be
g_{AB}=\int d^3x\sum_M \frac{\d \phi^M(\vec{x},t)}{\d q^A}\frac{\d \phi^M(\vec{x},t)}{\d q^B}\;,
\ee
where $\phi^M$ stands for the classical fields in our model, on the classical solution. Our fields are $F_{\mu\nu},\tilde F_{\mu\nu}, \a,\b$ and 
$\phi^{\a\b,\mu\nu}$. The classical Hopfion solution is 
\bea
&&\a=2i(t+z)-1\;,\;\; \b=2(x-iy)\;,\;\; \tilde F_{\mu\nu}=F_{\mu\nu}\cr
&&F_{\mu\nu}-\frac{i}{2}\epsilon_{\mu\nu\rho\sigma}F^{\rho\sigma}=-\epsilon_{\mu\nu\rho\sigma}\d_\rho\a\d_\b\sigma.
\eea
But then, on the classical solution, we obtain 
\bea
&&\frac{\d F_{\mu\nu}}{\d a}=0=\frac{\d F_{\mu\nu}}{\d b}=\frac{\d \tilde F_{\mu\nu}}{\d a}=\frac{\d \tilde F_{\mu\nu}}{\d b}\cr
&& \frac{\d \a}{\d a}=1\;,\;\;\; \frac{\d \b}{\d b}=1\;,\;\;\; \frac{\d \a}{\d b}=0=\frac{\d \b}{\d a}\;,
\eea
which means that the metric is 
\be
g_{aa}=\int dV 1=V=g_{bb}\;,\;\;\; g_{ab}=0\;,
\ee
so the Hamiltonian for the collective coordinates is, even at the quantum level, 
\be
H=\frac{1}{V}\left(\frac{\d}{\d a}^2+\frac{\d}{\d b}^2\right)+...\;,
\ee
which is trivial, and with a vanishing prefactor. Note that in the case of a usual soliton, $\int dV \phi'^2$ is a finite energy, so this issue doesn't arise.

But we can trace the reason for this result to something more fundamental. The sum of two fluctuations, i.e., small solutions, is in general not a 
solution of our action anymore. Indeed, as we said, even two propagating waves in different directions (each of which is a solution), is not a 
solution anymore. But linearity (the sum of solutions is a solution, at least in perturbation theory)
was an implicit assumption in the general formalism of collective coordinate quantization, and the role of nonlinearities 
of the action is simply to modify the solution order by order in some expansion parameter like a velocity $\dot X_0(t)$. Also, we 
have implicitly assumed that we can add fluctuations to a large classical solution, and it can still be a solution, which is again not true. 

If we then, for instance, blindly make the usual replacement $a\rightarrow a(t)$ and $b\rightarrow b(t)$ in our action, we can find 
a term in the action proportional to $\dot a(t)^2$ and $\dot b(t)^2$, just that we don't have a solution anymore, and now it is not even possible 
to find one perturbatively. Moreover, we must find an infinite prefactor for these terms, proportional to the volume $V$, as before, which would mean that 
the theory is only consistent for $\dot a=0=\dot b$. 

That means that the collective coordinate quantization, the usual way to deal with semiclassical quantization in field theory, is simply not 
applicable in our case, due to its quasi-topological nature. We must therefore use other means, as explained in previous sections.

\section{Conclusions and discussion}

In this paper we have constructed a relativistic
Lagrangian formulation for Bateman's construction for null configurations that include the Hopfion and other
knotted solutions. We have also written a Hamiltonian formulation for the same. We used these two formulations to construction {\em in principle}
the quantization of the null subsystem of electromagnetism, via path integral in the Lagrangian case and Dirac quantization in the Hamiltonian case, 
but doing the explicit construction (calculating observables in the first case, and constructing the Dirac brackets in the second case) 
seems very complicated in both cases.   We have shown that a semiclassical quantization of the 
system, using the collective coordinates method, doesn't work (at least in the usual way) due to the quasi-topological nature of the system.
It is not clear if another version of semiclassical quantization (quantization of fluctuations) could work.

We have also used the symmetries of the null system to construct generalizations of the Bateman null construction to a null free massless complex 
scalar, null massless 2-form and null DBI scalar, and show how to find solutions in them. 

There are many open questions that are still awaiting  further study, in particular:
\begin{itemize}
\item
In a similar manner to the Lagrangian formulation using Bateman's variables for the null electromagnetic theory one can probably write down similar formulations for the other null systems that we have mentioned. It might be that one can construct a Lagrangian for a general null system for which the systems discussed in this paper are special cases.
\item
An important question is to fully classify the knots associated with  all the  null configurations and to relate them to the standard mathematical classification of knots.
\item
We have in mind to perform explicit calculations of certain quantum properties of the null configurations. We have determined  several frameworks to achieve this goal but in the current paper we have not applied them to specific computations.  In particular the observations made here about problems in semi-classical quantization will be further studied.
\item
Probably the most challenging question is determining ways which will enable measuring the knotted null configuraions in the laboratory whether it is in the context of electromagnetic theory, hydrodynamics or any other physical system. 
\item
The application of special conformal transformations with imaginary parameters was shown in \cite{Hoyos:2015bxa} to be a powerful tool to construct topologically non-trivial solutions.  It is obvious that this approach has not been yet exhausted and deserves further exploration.  
\item
An interesting question is if the additional  systems addressed in this note  are relevant 
in some physical case, since we don't know of any fundamental massless scalar, DBI scalar or 2-form. Another would be to quantize these systems 
as well. 
\end{itemize}

\section*{Acknowledgements}
We would like to thank Carlos Hoyos, who took part in the early stages of this project and to  Daniel W.F. Alves, 
Manuel Array\'as  and Nilanjan Sircar for useful comments and discussions.
HN is grateful for the hospitality of the Department of Physics at Tel Aviv University, during which this work was started. 
This work was supported in part by a center of excellence supported by the Israel Science Foundation (grant number 1989/14), and by 
the US-Israel bi-national fund (BSF) grant number 2012383 and the Germany Israel bi-national fund GIF grant number I-244-303.7-2013. 
 J.S. would like to thank the theory group of Imperial College London and the Leverhulme trust for supporting his stay at Imperial 
 College where part of this work has been carried out.
The work of HN is supported in part by CNPq grant 304006/2016-5 and FAPESP grant 2014/18634-9. HN would also 
like to thank the ICTP-SAIFR for their support through FAPESP grant 2016/01343-7.

\bibliography{Hopfionpaper}
\bibliographystyle{utphys}

\end{document}